# Security management for backhaul-aware 5G-V2X

Vishal Sharma, Yongho Ko, Jiyoon Kim, Ilsun You
Information Security Engineering Department, Soonchunhyang University, Asan-si-31538, The Republic of Korea
vishal_sharma2012@hotmail.com, koyh0911@gmail.com, 74jykim@gmail.com, ilsunu@gmail.com

*Abstract*— Security is a primary concern for the networks aiming at the utilization of Cellular (C) services for connecting Vehicles to Everything (V2X). At present, C-V2X is observing a paradigm shift from Long Term Evolution (LTE) – Evolved Universal Terrestrial Radio Access Network (E-UTRAN) to Fifth Generation (5G) based functional architecture. However, security and credential management are still concerns to be resolved under 5G-V2X. A sizably voluminous number of key updates and non-availability of sub-functions at the edge cause adscititious overheads and decrement the performance while alarming the possibilities of variants of cyber attacks. In this paper, security management is studied as a principle of sustainability and its tradeoff is evaluated with the number of key-updates required to maintain an authenticated connection of a vehicle to the 5G-terminals keeping intact the security functions at the backhaul. A numerical study is presented to determine the claims and understand the proposed tradeoff.

## I. INTRODUCTION

With the availability of new radio access technologies, provisioning different types of services across the Cellular-Vehicle to Everything (C-V2X) is seen as the forefront of 5G networks [1] [2] [3]. This has been dominatingly termed as 5G-V2X where the core security and general functions are considered for facilitating the security and services for vehicles involved in the formation of the Vehicle to Vehicle (V2V), Vehicle to Infrastructure (V2I) and Vehicle to Pedestrian (V2P) networks as a part of V2X. Connecting fronthaul entities to the network and ensuring much of the operations at the edge require specific divisibility phenomenon which can simultaneously acquire as well as manage the backhaul operations [3] [4].

Most of the existing studies have presented this as a resource allocation problem [5] [6]. However, there is no evident study available which aims at providing a strategic management of security for 5G-V2X while considering the appropriate security for the backhaul formed between the Terminal (TM) and the hub as a part of the basic architecture. The initial reports on 5G have provided a specific security function architecture which can be integrated with the required base model as well as a security protocol to authenticate the devices involved in the transmissions [2] [7]. Initially, 5G-Authentication and Key Agreement (AKA) protocol and Extensible Authentication Protocol (EAP)-AKA prime are seen as competitive solutions for authenticating network entities [2]. However, the initial reports do not confirm the overheads associated with the periodic key updates, mobility management of vehicles, as well as the sustainability of V2X against a known set of cyber threats [8][9]. Periodic key updates undoubtedly enhance the security of a network, but this procedure, in the case of limited requirements, causes an uncertain burden on the entities and may result in high computational complexity as well as high operational cost [6] [10]. Thus, this article helps to understand the requirements of security management for V2X and also considers the possibilities of sub-dividing the 5G security functions to make it suitable for handling operations as well as authentication procedures of V2X at a high rate of sustainability with lesser key-updates.

The remaining part of this article is organized as follows: Section II presents the problem statement and highlights our contribution. Section III discusses the system model. The proposed approach is presented in Section IV. Section V covers the numerical evaluations. Finally, Section VI concludes the article.

## II. PROBLEM STATEMENT AND OUR CONTRIBUTION

Managing security for backhaul-aware 5G-V2X is tedious as it depends on the architectural deployment and 5G function-mappings to the underlaid network. Moreover, with a key focus on edge-initiation, bringing security-aspects near to users raise concerns of trustworthy security facilitations, especially when the involved entities are vehicles. With varying mobility and high dynamics in V2V, V2I and V2P formations, management of security needs an efficient solution which can protect the network under major attack scenarios. In this paper, the key exchange passes between the vehicles and 5G-security functions, which are deployed on the core, hub and the switches, are considered. The article also considers the backhaul between the TM and the hub for management as a part of the 5G deployment. The primary objective is to identify the time slots when the keys should be updated in order to protect the network against known cyber threats. Alongside, identification of fail-safe points up to which the network can be operated with the current keys is to be determined. Credential management, and certifying the key-hierarchy principles or enabling edge-initiated security are further considered as a part of the derived optimization problem. It should be considered with an extreme importance that updating keys is a crucial part of the network which has to deal with a large number of applications. Network with a large number of key-updates can be secured up to much extent but at the cost of performance. Thus, this tradeoff needs to be balanced for attaining an efficient backhaul-aware 5G-V2X. Based on the solutions presented in this article, the core contributions can be marked as follows:

- Dual security management solution is presented that considers security through long-range and short-range authentications.

- A fail-safe point is identified as a part of the optimization solution, until which, the network can be operated without changing the derived keys.
- A new hierarchy principle is presented for managing key derivations from the 5G security functions.

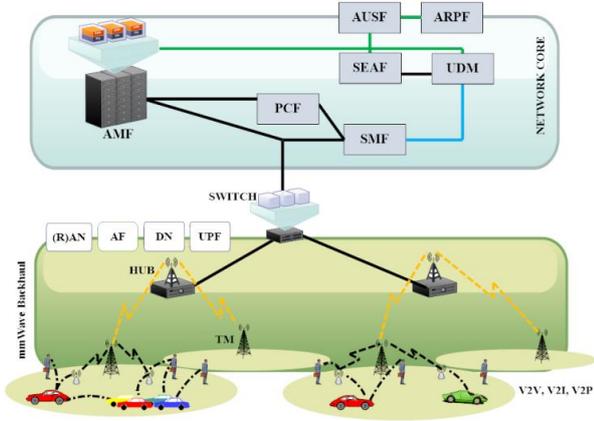

Fig.1 An illustration of 5G-V2X with mmWave Backhaul between the Hub and the terminal based on 5G security and general network functions.

### III. SYSTEM MODEL

The network comprises 5G security functions which are mapped across the core, TMs, switches, User Equipment (UE), and hubs, as shown in Fig.1. The UEs involve the vehicles and general users that form V2V, V2I and V2P links during their operations. The core involves the Access and Mobility Management Function (AMF), which operates together with the Session Management Function (SMF) and Policy Control Function (PCF), and as described in the initial TS by 3GPP [2], the Security Anchor Function (SEAF) is placed in their periphery which is located deep in the network without any knowledge to the edge as well as vehicles. As it is tedious to consider such a deployment for supporting edge-initiated security as well as authentication for V2X, the proposed part modifies it and presents a new disintegrated structure to make the 5G-drafted version suitable for V2X. SEAF keys and authentication are driven by the Authentication Server Function (AUSF) and the Authentication Credential Repository and Processing Function (ARPF) [2].

The mmWave backhaul supports the Unified Data Management (UDM), Application Function (AF), (Radio) Access Network ((R)-AN), and User Plane Function (UPF). In accordance with the problem statement, the issue is with the security management that involves the V2X authentication and TM authentication with the hub along with the placement of major details at the edge without costing the performance. This article considers security to be directly proportional to the number of key-updates performed in the V2X setup. However, with an excessive number of updates, the network induces certain overheads, which slow down the operations and increase the burden of credential management. To resolve this, at first, a timing problem is formulated. According to which, let $t$ be the time taken by an adversary to launch an attack, out of which $t'$ be the minimum time for which the keys should not be changed.

Considering this, the key-utilization time, $t_u$, must be identified, such that $t_u < t'$ ($<=t$), for which the currently allocated keys can be safe from known attacks. Moreover, intermediate-key updates, $U_k$, should be minimized while maximizing the overall sustainability, $S_N$, of the network. Here, $S_N$ is derived as a function of induced overheads by using IPAT formulations [11], such that

$$S_N = \frac{nU_k}{D.P.Q}, \quad (1)$$

where $D (\leq N) = \int_{r_1}^{r_2} C(x, y)dx$ is the number of cars in the periphery ($r$) of a particular TM for a given density function $C(x,y)$. $P$ is the probability of loss in connectivity; $Q$ is the number of passes between the entities, $n$ is the sustainability balancing constant depicting the inverse of number of hops between the vehicles and the entity dealing with the particular request generated for a 5G security function (key or service requests), $N$ is the end devices (vehicles, UEs, or users), and $E$ is the overall involved entities. Now, based on the above-discussed security issues, the optimization problem can be formulated as:

$$\max(S_N) \, \forall E, \forall N, \quad (2)$$

s.t.

$$\max(t_u), \forall N,$$

$$\underbrace{\min(U_N)}_{\text{in tradeoff with } S_N}, \text{ and } U_N \geq U'_N,$$

$$0 < D \leq N,$$

$$0 < \frac{n^{-1}(n^{-1}-1)}{2} \leq \frac{E(E-1)}{2}, n^{-1} \neq E,$$

$$\min(t_u - t'). \quad (3)$$

Here, $U'_N$ is the mandatory key-updates below which it is difficult to evaluate the network for security and perform any tasks, such as mobility management, re-authentication.

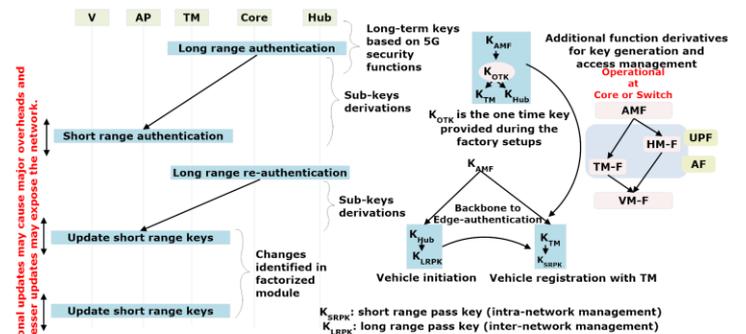

Fig.2 An overview of the security considerations, key generations, long and short-range authentication and additional functions derived for managing sub-keys.

### IV. PROPOSED APPROACH

The proposed approach establishes a security management framework which operates as a dual security driving system based on two modes for authentication-long range authentication and short-range authentication. Both these help to provide a strong authentication for V2X while ensuring the security of backhaul between the TM and the hub. The proposed approach maintains a plan for authentication via key exchanges that are performed by utilizing the deployed 5G security functions. The long-range authentications are performed for the backhaul whereas the short-range authentications are performed for the edge-initiated network formations. The initial mode is dependent on the derivations of keys from the core, which is a part of long-range authentication and sub-division of initial keys support the procedures for short-range authentication. The long-range authentication is operated over a powerful mmWave communications, which can be facilitated by existing encryption algorithms and can be motivated to update the keys once the devices are deployed or reconfigured. However, the major issue prevails at the edge-initiated authentication for the vehicles, which under high dynamics, causes an additional burden of key-regenerations as well as credential management. To further resolve this issue, a factorized module is considered in the short-range authentication, which evaluates the speed (S), location (L), last updates for keys ($U_T$), shared sessions ($A_S$), refreshing rate of keys ($F_R$), total keys ($T_K$), zone traversals ($Z_T$), and associatively ($V_A$) of a vehicle to generate new keys or continue with the existing keys while maintaining the conditions stated in (2). An overview of the proposed approach with key dependencies is presented in Fig.2 with the description as follows:

- The AMF function is sub-divided into the terminal function (TM-F) and Hub function (HM-F) that derive their keys from $K_{OTK}$, which is derived from $K_{AMF}$.
- The derived keys are used for initial long-range authentications and short-range authentications. The sub-derived keys are used for authenticating vehicles' fronthaul to the edge and the backhaul between the TM and the hub.
- The sub-derivative functions help to support the intra- and inter-mode of secure communications in V2X by deriving their core operations from the 5G security functions.

At present, the details of internal procedures for the proposed framework and authentication mechanisms are omitted from the article and prime focus has been given to understand the sustainability of the backhaul-aware 5G-V2X under the given constraints of key updates. These can further be observed from the following results:

**Lemma-1:** $S_N$ is divergent if the initial information on the key generation is unavailable. However, in the case of known timestamps, the network sustainability can be modeled over the available key updates and vehicle movement, such that for instances $t_1$ and $t_2$:

$$S_N = \frac{\alpha^2}{2\beta N \left(1 - \frac{n-1}{E}\right)^N Q} \left(Ei\left(\frac{\beta - \alpha}{t_1}\right) - Ei\left(\frac{\beta - \alpha}{t_2}\right)\right),$$
(4)

at $t_2 - t_1 > 0$, $E - n^{-1} > 0$, $\beta - \alpha > 0$, and $Ei$ is the exponential integral.

**Proof:** Considering (1) to be following Poisson distribution for vehicles approaching at a rate of $\beta$ vehicles per unit time and operating with $\alpha$ number of key-updates per unit time, $U_k$ can be given by $e^{-\frac{\alpha}{t}} \frac{\left(\frac{\alpha}{t}\right)^X}{X!}$, (X=2), as only 2 keys are used for authentication (long range and short range keys) and D can be written as $e^{-\frac{\beta}{t}} \frac{\left(\frac{\beta}{t}\right)^{X'}}{X'!}$, (X'=1), as each vehicle maintains its connectivity with one source from the network. Now, considering the involved entities in the network for authenticating a vehicle, and using the model in [12], the probability of no connectivity (P) can be given as $\left(1 - \frac{n-1}{E}\right)^N$. The number of passes remains at the discretion of the used protocol and is constant for this evaluation. By using these values in (1) and under fixed interval, the observed equation can be written as:

$$S_N = \frac{1}{N\left(1 - \frac{n-1}{E}\right)^N Q} \int_{t_1}^{t_2} \frac{e^{-\frac{\alpha}{t}} \frac{\left(\frac{\alpha}{t}\right)^2}{2!}}{e^{-\frac{\beta}{t}} \frac{\left(\frac{\beta}{t}\right)^1}{1!}} dt .$$
(5)

On solving, at $t_2-t_1>0$, $E-n^{-1}>0$, $\beta - \alpha>0$, the observation is

$$S_N = \frac{\alpha^2}{2\beta N \left(1 - \frac{n-1}{E}\right)^N Q} \left(Ei\left(\frac{\beta - \alpha}{t_1}\right) - Ei\left(\frac{\beta - \alpha}{t_2}\right)\right),$$
(6)

which is the desired output.

**Lemma-2:** For extreme large rates of key exchanges and vehicle dynamics, $S_N$ is convergent to a linear function under asymptotic observations. This can be further used to identify fail-safe points with high accuracy up to which the network can be operated without much overheads and security breaches.

**Proof:** In continuation from Lemma-1, the result in (6) can be asymptotically analyzed for evaluating the behavior of the curve determining the sustainability of the V2X under given constraints. According to which, if $\alpha$ and $\beta$ increases to a larger value, the (6) reduces to a linear function, such that $S_N$ can be evaluated as a function comprising two metrics, i.e. $S_N = f\left(\frac{\alpha}{t}, \frac{\beta}{t}\right)$; and by following strict principles and limits, it can be given as $S_N = \frac{\alpha}{\beta}$. Moreover, in such a case, the behavior is not integrally divergent and the network can be evaluated with linear timestamping. Now, considering these observations, the network fail-safe points up to which there is no need to update or refresh the keys can be evaluated as:

$$F_S = \begin{cases} t \text{ at } S_N \geq S_N^{TH}, \text{if } t_1 \neq 0, \text{and known} \\ t \text{ at } M_O \leq M_O^{TH}, \text{otherwise} \end{cases}.$$
(7)

Here, $M_O$ is the message overhead evaluated as

$$M_O = \frac{O_S(1-P)}{E.P},$$
(8)

where $O_S$ is said to be increasing exponentially. This is due to the reason that the key updates and vehicles follow the Poisson distribution with randomness observable from Lemma-1, based on which, $O_s$ can be modeled as $O_b\left(1 - \frac{\alpha}{t}\right)^t$, where $O_b$ is the overheads for initial authentication. This can be predicted between the instances, as shown in Lemma-1, only at a fixed number of updates ($\alpha' = \frac{\alpha}{t}$), such that

$$O_S = \frac{O_b \left(\frac{n-1}{E}\right)^N}{E\left(1 - \frac{n-1}{E}\right)^N} \cdot \frac{\ln(1-\alpha')^{t2} - \ln(1-\alpha')^{t1}}{\ln(1-\alpha')}.$$
(9)

These observations are required to determine the exactness of updates for the available values of sustainability as well as message overheads. The thresholds are identified based on the data pool available for the time required to launch an attack particularly on the protocol used for the authentication of vehicles to everything or TM to the hub.

**Table 1. Parameter Settings**

| Metrics | Values | Metrics | Values |
|---|---|---|---|
| β | 2-10 (step 2) (A1-A5) | $n^{-1}$ | 5 |
| α | β/2 | Q | 1-5 |
| N | 10 | $t_1$ | 5 s |
| E | 1-10 | $t_2$ | 105 s |

## V. NUMERICAL EVALUATIONS

Security of backhaul-aware 5G-V2X cannot be quantified and it depends on the operations of the involved authentication protocol, which will be presented in our future reports. However, the timing to update keys and managing security based on the arrival rate of vehicles is quantifiable and it can be observed in the form of sustainability as expressed in the earlier part of this article. To further understand the impact of metrics on the sustainability and the location of fail-safe points, a numerical case study is conducted by using metrics defined in Table 1. The values are based on a general setup described in the initial parts of this article and other metrics are assumed within the constraints of Lemma-1 and Lemma-2. The result is recorded for $S_N$, as shown in Fig.3.

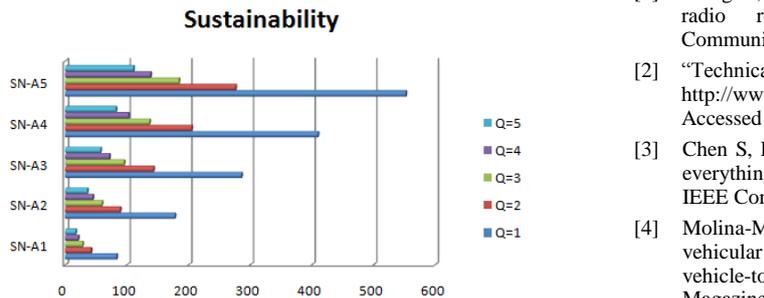

Fig. 3 Network sustainability w.r.t. the number of protocol passes at different rates for incoming vehicles and the number of key updates.

The results suggest that with the proposed strategy, it is feasible to effectively trace the network activity and manage its security by keeping a check on unnecessary key updates. By this, significant overheads can be reduced from the network and the security can be provided at the edge while keeping a close associatively with the network-backhaul. It is observed that the number of passes and key-updates used for authentication pose a significant impact on the performance as well as the security of the network. Numerically, the results vary between 53.1% and 84.9% compared to initial observed output at constant arrival and update rates for vehicles and keys, respectively, with a varying value of Q. It is evident that the sustainability of the network can be enhanced even with an increase in the number of vehicles or key-updates by reducing the number of passes between the vehicle and the infrastructure. Thus, it is important to carefully select the authentication protocol and it must not cause excessive signaling overheads. For this, the proposed approach leverages on the divisibility of 5G security functions based on the requirements at the specific edge leading to the formations of TM-F and HM-F, which help to effectively manage the security in a backhaul-aware 5G-V2X.

## VI. CONCLUSION

In this article, a tradeoff between the sustainability and the number of key-updates is managed in a backhaul-aware 5G-V2X. A security management framework is proposed which considers security through long- and short-range authentications. Analytical evaluations are presented to study the impact of key-exchanges, the arrival rate of vehicles and the number of authentication passes on the sustainability of the network. Following which a fail-safe point is identified as a part of the optimization solution. To generalize, new functions are derived to handle the conceptualization of the proposed solution.

This is an incremental article and further information on the authentication procedures, key-exchanges, and operational details will be presented in our future reports.

## ACKNOWLEDGMENT

This work was supported by 'The Cross-Ministry Giga KOREA Project' grant funded by the Korea government (MSIT) (No.GK18N0600, Development of 20Gbps P2MP wireless backhaul for 5G convergence service).